\documentclass[twocolumn,prb,showpacs]{revtex4}
\usepackage{graphicx}
\usepackage{dcolumn}
\usepackage{bm}
\usepackage{amsmath}

\begin{document}

\preprint{}
\title{Controllable valley and spin transports in ferromagnetic silicene junctions}
\author{Takehito Yokoyama}
\affiliation{Department of Physics, Tokyo Institute of Technology, Tokyo 152-8551,
Japan 
%\\ $^2$Department of Physics, Tokyo Metropolitan University, Hachioji, Tokyo 192-0397, Japan 
}
\date{\today}

\begin{abstract}
We investigate charge, valley, and spin transports in normal/ferromagnetic/normal silicene junction. We show that the charge, valley,  and spin conductances in this junction oscillate with the length of the ferromagnetic silicene. It is also found that the current through this junction is valley and spin polarized due to the coupling between valley and spin degrees of freedom, and the valley and spin polarizations can be tuned by local application of a gate voltage. In particular, we find \textit{fully} valley and spin polarized current by applying the electric field. 
We also obtain the condition for observing the fully valley and spin polarized current.

\end{abstract}

\maketitle

%\affiliation{$^1$Department of Physics, Tokyo Institute of Technology, 2-12-1 Ookayama, Meguro-ku, Tokyo 152-8551, Japan \\

Silicene is a monolayer of slilicon atoms on a two dimensional honeycomb lattice.\cite{Takeda}  Recently, this material has been successfully synthesized.\cite{Lalmi,Padova,Padova2,Vogt,Lin,Fleurence}
Due to the lattice structure, electrons in silicene obey the Dirac equation around the $K$ and $K'$ points at low energy.\cite{Liu,Liu2} Silicene has a large spin-orbit gap compared to graphene, and due to the buckled structure, the mass of the Dirac electrons are controllable by external electric field.\cite{Ezawa,Ezawa2} With this property, it has been predicted that there occurs a topological transition between band and topological insulators by applying electric field. \cite{Ezawa,Ezawa2}
Also, transport property in silicene based pn junction has been investigated. \cite{Yamakage}
In this paper, we study valley and spin transports in silicene junction. 

Valleytronics aims to control valley transport by electric means and vice versa and has developed in graphene\cite{Rycerz,Xiao,Akhmerov}. In graphene nanoribbons with a zigzag edge, 
valley filter and valley valve effect have been proposed.\cite{Rycerz,Akhmerov} These originate from intervalley scattering by a potential step and are hence controllable by local application of a gate voltage. 
Also, graphene exhibits gate-controlled carrier conduction, high field-effect mobilities, and a small spin-orbit interaction.\cite{Kane,Hernando} Therefore, graphene offers a good arena for observing spintronics effects. \cite{Son,Kan,Yazyev,Haugen,Tombros,Yokoyama,Yokoyama2}
In graphene with ferromagnet deposited on the top, spin transport controlled by a gate electrode has been predicted. \cite{Haugen,Yokoyama,Yokoyama2}
There are also attempts to use pseudospin degrees of freedom in graphene to obtain new functionalities. \cite{Jose,Xia,Majidi}

In this paper, we study charge, valley, and spin transports in normal/ferromagnetic/normal silicene junction. We show that the charge, valley,  and spin conductances in this junction oscillate with the length of the ferromagnetic silicene. It is also found that the current through this junction is valley and spin polarized due to the coupling between valley and spin degrees of freedom, and the valley and spin polarizations are tunable by local application of a gate voltage. In particular, we find \textit{fully} valley and spin polarized current by applying the electric field. 
We also obtain the condition for observing the fully valley and spin polarized current.

%%%%%%%%%%%%%%%%%%%%% Formulation
%\vspace{2cm}

\begin{figure}[tbp]
\begin{center}
\scalebox{0.8}{
\includegraphics[width=11.0cm,clip]{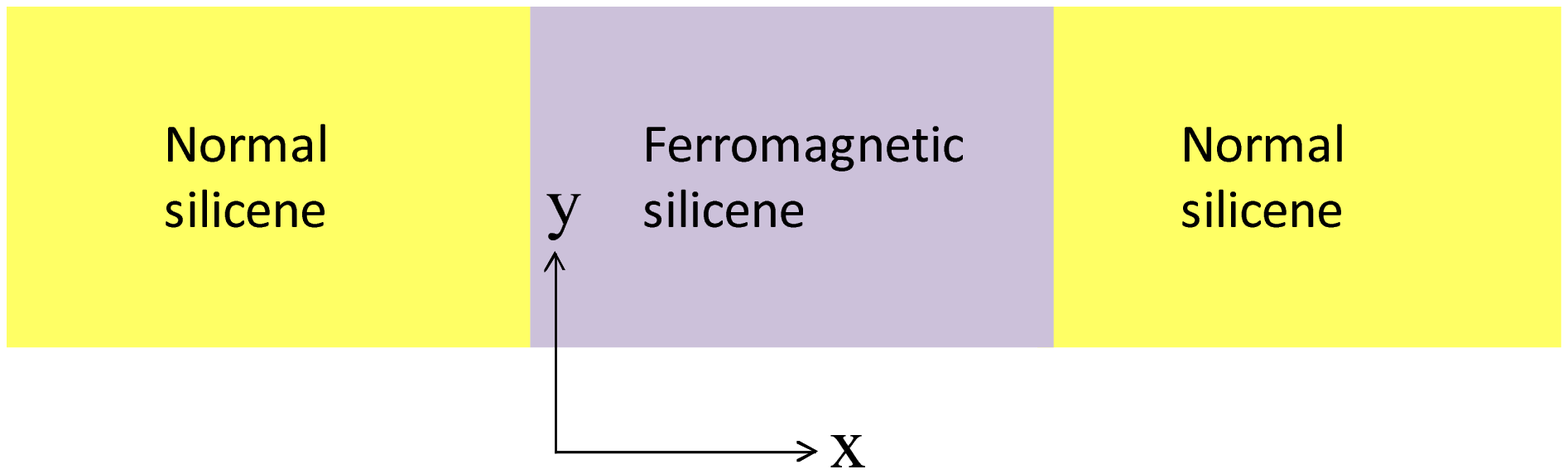}
}
\end{center}
\caption{ (Color online) Schematic picture of the model. }
\label{fig1}
\end{figure}

%\vspace{2cm}

We consider a normal/ferromagnetic/normal silicene junction as shown in Fig. \ref{fig1}.
The Hamiltonian of the ferromagnetic silicene is given by \cite{Liu,Liu2,Ezawa,Ezawa2}
\begin{eqnarray}
H = \hbar v_F (k_x \tau _x  - \eta k_y \tau _y ) - \Delta _{\eta \sigma } \tau _z  - \sigma h
\end{eqnarray}
with $\Delta _{\eta \sigma }  = \eta \sigma \Delta _{so}  - \Delta _z $. $\tau$ is the Pauli matrix in sublattice pseudospin space. $\Delta _{so}$ denotes the spin-orbit coupling. $\Delta _z$ is the onsite potential difference between $A$ and $B$ sublattices, which is tunable by an electric field applied perpendicular to the plane. $h$ is the exchange field in the ferromagnetic region. $\eta  =  \pm 1$ corresponds to the $K$ and $K'$ points. $\sigma  =  \pm 1$ denotes the spin indices. 
The large value of $\Delta _{so}= 3.9$ meV in silicene \cite{Liu2} leads to a cross correlation between valley and spin degrees of freedom, which is a clear distinction from graphene.
Magnetism could be induced in silicene by the magnetic proximity effect with a magnetic insulator EuO as proposed for graphene.\cite{Haugen} 
In the normal region, we set $\Delta _z=h=0$. 
Thus, the gate electrode is attached on the ferromagnetic segment. The eigenvalues of the Hamiltonian in the normal and ferromagnetic silicene are given by 
\begin{eqnarray}
E =  \pm \sqrt {(\hbar v_F k)^2  + (\Delta _N )^2 } \nonumber \\
 =  \pm \sqrt {(\hbar v_F k')^2  + (\Delta _F )^2 }  - \sigma h
\end{eqnarray}
with $\Delta _N  = \eta \sigma \Delta _{so}$ and $\Delta _F  = \eta \sigma \Delta _{so}  - \Delta _z$. $k$ and $k'$ are momenta in the normal and the ferromagnetic regions, respectively. 
Let $x$-axis perpendicular to the inteface and assume the translational invariance along the interface, $y$-axis.  The interfaces between the normal and the ferromagnetic silicene are located at $x=0$ and $x=L$ where $L$ is the length of the ferromagnetic silicene. Then, the wavefunctions for valley $\eta$ and spin $\sigma$ in each region can be written as 
\begin{widetext}
\begin{eqnarray}
 \psi (x < 0) = \frac{1}{{\sqrt {2EE_N } }}e^{ik_x x} \left( {\begin{array}{*{20}c}
   {\hbar v_F k_ +  }  \\
   {E_N }  \\
\end{array}} \right) + \frac{{r_{\eta ,\sigma } }}{{\sqrt {2EE_N } }}e^{ - ik_x x} \left( {\begin{array}{*{20}c}
   { - \hbar v_F k_ -  }  \\
   {E_N }  \\
\end{array}} \right), \\ 
 \psi (0 < x < L) = a_{\eta ,\sigma } e^{ik'_x x} \left( {\begin{array}{*{20}c}
   {\hbar v_F k'_ +  }  \\
   {E_F }  \\
\end{array}} \right) + b_{\eta ,\sigma } e^{ - ik'_x x} \left( {\begin{array}{*{20}c}
   { - \hbar v_F k'_ -  }  \\
   {E_F }  \\
\end{array}} \right), \\ 
 \psi (L < x) = \frac{{t_{\eta ,\sigma } }}{{\sqrt {2EE_N } }}e^{ik_x x} \left( {\begin{array}{*{20}c}
   {\hbar v_F k_ +  }  \\
   {E_N }  \\
\end{array}} \right)
\end{eqnarray}
\end{widetext}
with $\hbar v_F k'_x  = \sqrt {(E + \sigma h)^2  - (\Delta _F )^2  - (\hbar v_F k_y )^2 } $, $ E_N  = E + \Delta _N ,E_F  = E + \sigma h + \Delta _F$, and $k_ \pm ^{(\prime)}  = k_x^{(\prime)}  \pm i\eta k_y$. 
Here, ${r_{\eta ,\sigma } }$ and ${t_{\eta ,\sigma } }$ are reflection and transmission coefficients, respectively.
By matching the wavefunctions at the interfaces, we obtain the transmission coefficient: 
\begin{eqnarray}
 t_{\eta ,\sigma }  = 4k_x k'_x E_N E_F e^{ - ik_x L} /A,\quad  \\ 
  A = (\alpha ^{ - 1}  - \alpha )k^2 E_F^2  + (\alpha ^{ - 1}  - \alpha )(k')^2 E_N^2  \nonumber \\ 
  + E_N E_F \left[ {k_ +  (\alpha ^{ - 1} k'_ +   + \alpha k'_ -  ) + k_ -  (\alpha ^{ - 1} k'_ -   + \alpha k'_ +  )} \right]  
\end{eqnarray}
with $\alpha  = e^{ik'_x L}$. 

By setting $k_x  = k\cos \phi$ and $k_y  = k\sin \phi$,  we define normalized valley and spin resolved conductance: 
\begin{eqnarray}
G_{\eta \sigma }  = \frac{1}{2}\int_{ - \pi /2}^{\pi /2} {\left| {t_{\eta, \sigma } } \right|^2 \cos \phi d\phi } .
\end{eqnarray}
The valley resolved conductance is defined as $G_{K^{(')}}  = (G_{K^{(')} \uparrow }  + G_{K^{(')} \downarrow } )/2$. 
We also introduce charge conductance, $G_c$, and valley and spin polarization, $G_v$ and $G_s$:
\begin{eqnarray}
G_c  = G_K  + G_{K'} , \\ 
G_v  = \frac{{G_K  - G_{K'} }}{{G_K  + G_{K'} }}, \\ 
G_s  = \frac{{G_{K \uparrow }  + G_{K' \uparrow }  - G_{K \downarrow }  - G_{K' \downarrow } }}{{G_K  + G_{K'} }} .
\end{eqnarray}

In the following, we fix $h$ and $\Delta_{so}$ as $h/E=0.3$ (except for Fig. \ref{fig6}) and $\Delta_{so}/E=0.5$.
We consider finite chemical potential by doping or gating in silicene.

\begin{figure}[tbp]
\begin{center}
\scalebox{0.8}{
\includegraphics[width=11.0cm,clip]{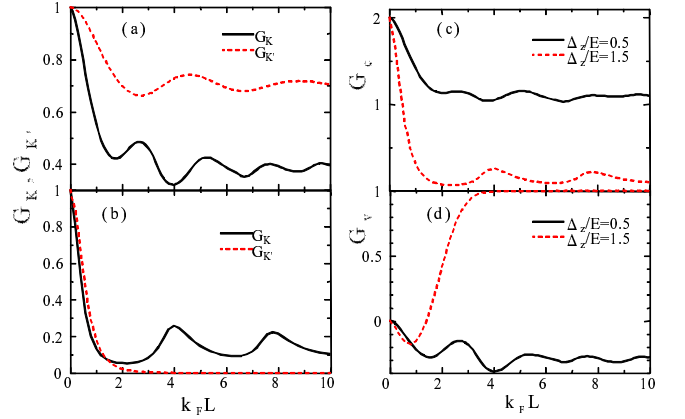}
}
\end{center}
\caption{ (Color online) (a) Valley resolved conductance $G_{K^{(')}}$ for $\Delta_z/E=0.5$ as a function of $L$. (b) Valley resolved conductance $G_{K^{(')}}$ for $\Delta_z/E=1.5$  as a function of $L$. (c) Charge conductance $G_c$ as a function of $L$. (d) Valley polarization $G_v$ as a function of $L$. }
\label{fig2}
\end{figure}

\begin{figure}[tbp]
\begin{center}
\scalebox{0.8}{
\includegraphics[width=11.0cm,clip]{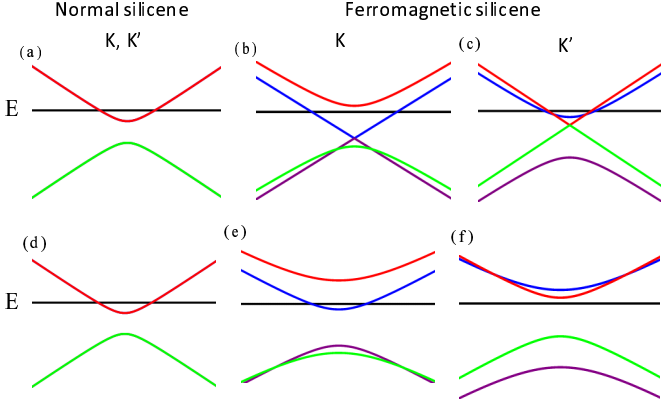}
}
\end{center}
\caption{ (Color online) Band structures near the $K$ and $K'$ points in normal silicene, (a) and (d), and ferromagnetic silicene, (b), (c), (e), and (f). The horizontal line denotes the Fermi energy. The upper panels correspond to the bands for small $\Delta_z$ (e.g., $\Delta_z/E=0.5$) while lower panels correspond to those for large $\Delta_z$ (e.g., $\Delta_z/E=1.5$).}
\label{fig3}
\end{figure}

\begin{figure}[tbp]
\begin{center}
\scalebox{0.8}{
\includegraphics[width=11.0cm,clip]{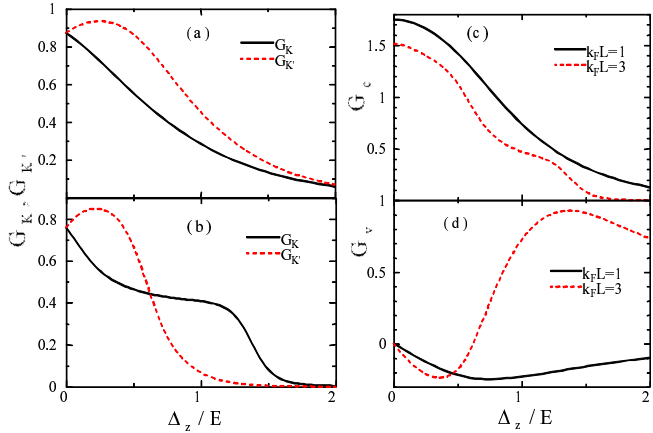}
}
\end{center}
\caption{ (Color online) (a) Valley resolved conductance $G_{K^{(')}}$ for $k_F L=1$ as a function of $\Delta_z$. (b) Valley resolved conductance for $k_F L=3$ as a function of $\Delta_z$. (c) Charge conductance $G_c$ as a function of $\Delta_z$. (d) Valley polarization $G_v$ as a function of $\Delta_z$. }
\label{fig4}
\end{figure}

\begin{figure}[tbp]
\begin{center}
\scalebox{0.8}{
\includegraphics[width=11.0cm,clip]{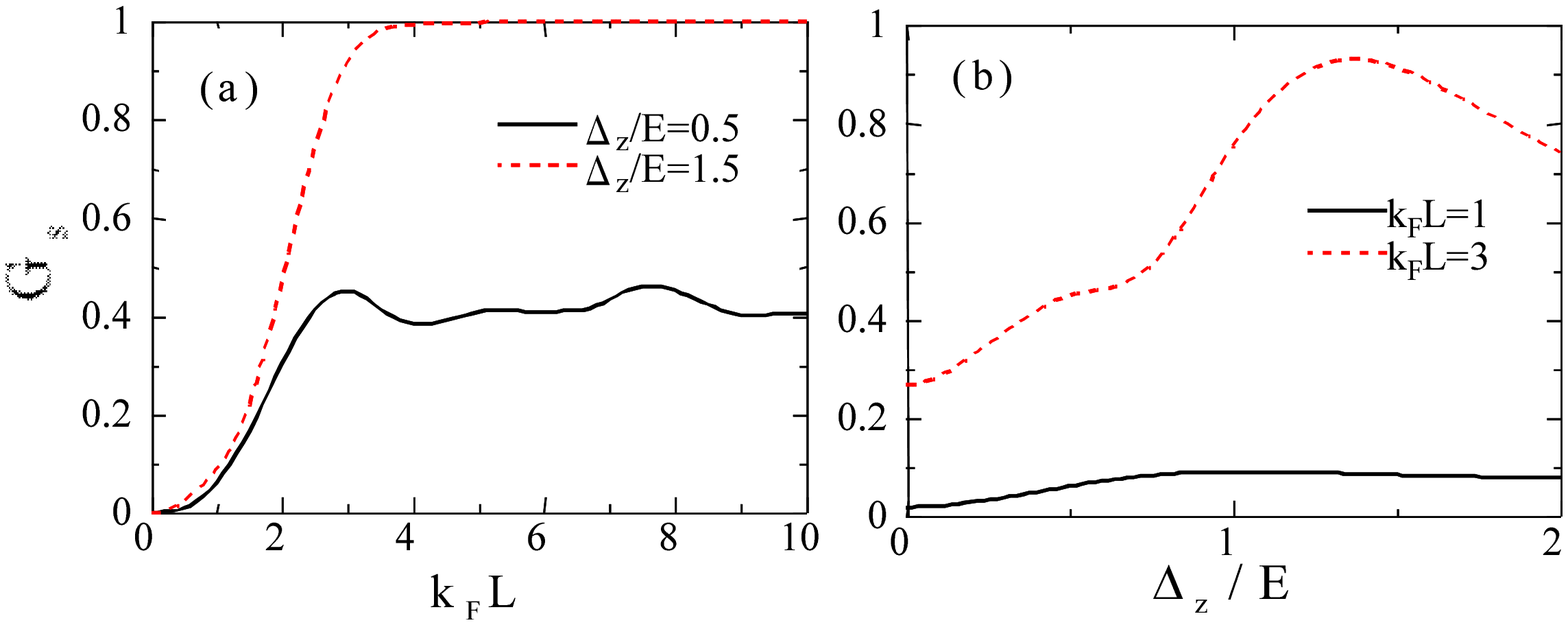}
}
\end{center}
\caption{ (Color online) Spin polarization $G_s$ as functions of (a) $L$ and (b) $\Delta_z$. }
\label{fig5}
\end{figure}

\begin{figure}[tbp]
\begin{center}
\scalebox{0.8}{
\includegraphics[width=8.50cm,clip]{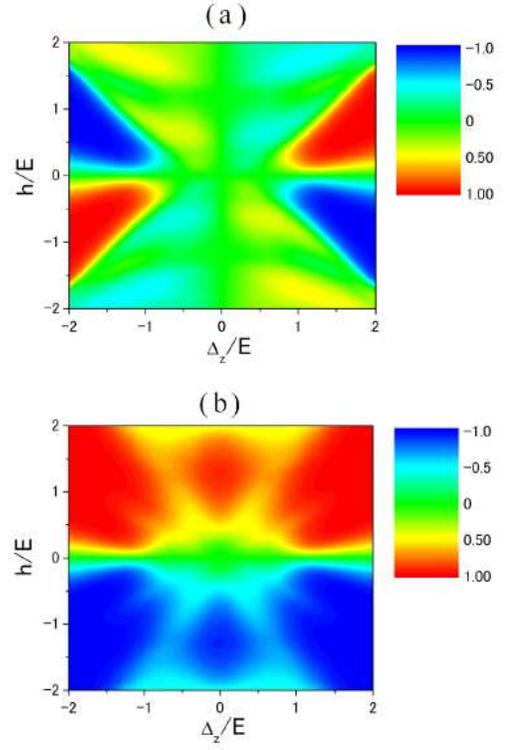}
}
\end{center}
\caption{ (Color online) (a) Valley polarization $G_v$ as functions of $\Delta_z$ and $h$. (b) Spin polarization $G_s$ as functions of $\Delta_z$ and $h$. }
\label{fig6}
\end{figure}

Figure \ref{fig2} displays valley resolved conductance $G_K$ and $G_{K'}$ for (a) $\Delta_z/E=0.5$ and (b) $\Delta_z/E=1.5$, (c) charge conductance $G_c$, and (d) valley polarization $G_v$ as a function of the length of the ferrmagnetic region $L$. Here, $k_F$ is defined as $k_F  = E/(\hbar v_F )$. For $\Delta_z/E=0.5$, both $G_K$ and $G_{K'}$ decays with $L$ in an oscillatory way. The oscillation is a reminiscent of the anomalous tunneling of massless Dirac fermion (Klein tunneling).\cite{Katsnelson} For $\Delta_z/E=1.5$, $G_K$ behaves similarly but $G_{K'}$ is strongly suppressed. Charge conductance $G_c$ depicted in Fig. \ref{fig2}(c) also shows an oscillatory dependence on $L$. The valley polarization $G_v$ for $\Delta_z/E=0.5$ depends on $L$ in an oscillatory fashion as shown in Fig. \ref{fig2}(d). The $G_v$ for $\Delta_z/E=1.5$  reach unity for  large $k_F L$, namely, \textit{the charge current is fully valley polarized.} This is because in the regime, the $G_{K'}$ becomes almost zero.

These characteristics can be understood by the band structures in each region as shown in  Fig. \ref{fig3}.  
Figure \ref{fig3} shows the band structures near the $K$ and $K'$ points in normal silicene, (a) and (d), and ferromagnetic silicene, (b), (c), (e), and (f). The horizontal line denotes the Fermi energy. The upper panels correspond to the bands for small $\Delta_z$ (e.g., $\Delta_z/E=0.5$) while lower panels correspond to those for large $\Delta_z$ (e.g., $\Delta_z/E=1.5$).  
For small $\Delta_z$, the band gap is small and hence the Fermi level crosses the bands near both $K$ and $K'$ points in the ferromagnetic region as shown in  Figs. \ref{fig3} (b) and (c). Thus, electrons near the $K$ and $K'$ points contribute to the current. Since electrons experience different ``potentials" near the $K$ and $K'$ points, the periods of the oscillations in $G_K$ and $G_{K'}$ in Fig. \ref{fig2} (a) are different. Around the $K$ point, the Fermi level crosses one band while near the $K'$ point, the Fermi level crosses two bands as shown in  Figs. \ref{fig3} (b) and (c). Hence, the current stemming from the $K'$ point is larger than that from the  $K$ point in this case.
For large $\Delta_z$, the Fermi level is located inside the band gap near the $K'$ point as shown in  Fig. \ref{fig3} (f). Then, the current is entirely carried by the electrons near the $K$ point --\textit{fully valley polarized current}. 
It should be noted that $\Delta_z$ can be tuned by the gate electrode. Therefore, we find a highly controllable valley transport in this junction.
We also note that in the absence of the exchange field, the charge conductance has equal contributions from the $K$ and $K'$ points.\cite{Yamakage} 

The condition to realize fully valley polarized transport can be obtained as follows. To locate the Fermi level $E (> \Delta_{so})$ within the band gap at the $K'$ point ($\eta  =  - 1$), $- \left| {\sigma \Delta _{so}  + \Delta _z } \right| - \sigma h < \Delta _{so}  < \left| {\sigma \Delta _{so}  + \Delta _z } \right| - \sigma h$ should be satisfied. Therefore, we obtain the condition necessary for the fully valley polarized transport as
\begin{eqnarray}
\Delta _z  > \max (h, \Delta _{so} , 2\Delta _{so}  - h) .
\end{eqnarray}

Figure \ref{fig4} illustrates (a) valley resolved conductance for $k_F L=1$, (b) valley resolved conductance for $k_F L=3$, (c) charge conductance, and  (d) valley polarization as a function of $\Delta_z$.
If the length of the ferromagnetic silecene is small, then, even if the Fermi level is located in the band gap near the $K'$ point, the current can flow due to the tunneling effect as shown in Fig. \ref{fig4} (a). For thick ferromagnetic region, with increasing  $\Delta_z$, current coming from the $K'$ point strongly decreases. Then, $G_K$ gives a dominant contribution to the current as  seen from Fig. \ref{fig4} (b). 
The charge conductance monotonically decreases with $\Delta_z$ as shown in Fig. \ref{fig4} (c).  Since for large $\Delta_z$, current is mostly carried by electrons around the $K$ point,  the valley polarization for $k_F L=3$ is strongly enhanced for large $\Delta_z$ as shown in Fig. \ref{fig4} (d).

We show the spin polarization $G_s$ as functions of (a) $L$ and (b) $\Delta_z$ in Fig. \ref{fig5}. Since the bands in the ferromagnetic region are also spin-split in a different way, depending on the $K$ and $K'$ points,\cite{Ezawa} we find that the current flowing through the junction is also spin polarized. 
For large $\Delta_z$, only single band at the $K$ point contributes to the current as shown in Fig. \ref{fig3}. Consequently, the current is \textit{fully} spin polarized as shown in Fig. \ref{fig5} (a). We again notice that  $\Delta_z$ is a tunable parameter. Hence, the spin current is also controllable in this setup as shown in Fig. \ref{fig5} (b). 
Combined with  Fig. \ref{fig2} (d), we find valley and spin polarized current  by local application of a gate voltage.

Figure \ref{fig6} shows (a) $G_v$ and (b) $G_s$ as functions of $\Delta_z$ and $h$ for $k_F L=3$. The $G_v$ is odd with respect to $\Delta_z$ and $h$. For large  $\Delta_z$, $G_v$ becomes large as we found in Fig. \ref{fig4}. However, for smaller $\Delta_z$, the magnitude of $G_v$ can be still $\sim 0.5$. 
The $G_s$ is odd in $h$ but even in $\Delta_z$. For large $h$, $G_s$ becomes large as expected. Even for small $h$, $G_s$ can be large for large $\Delta_z$.
From this figure, it is found that fully valley and spin polarized currents are realized for large $\Delta_z$ regime but relatively high polarizations ($\ge  0.5$) can be realized in a wide parameter region.

If we choose ferromagnetic silicene with $k_F L$ = 1 and $E=$10 meV, since $v_F \sim 5 \times 10^5$m/s, we need a junction length of order of 10 nm. 
For $\Delta_z \sim E$, the electric field applied perpendicular to the plane is about 34 meV/\AA $ $
 since the distance between the $A$ and $B$ sublattice planes is 0.46 \AA. 
%These values can be achieved by the present experimental techniques.
The exchange field induced in silicene due to the magnetic proxmity effect could be of the order of 1 meV for the magnetic insulator EuO.\cite{Haugen} 
Throughout the calculation, we assume the zero temperature limit. This assumption is justified for tempereture regime lower than $\Delta_{so}$, $\Delta_z$, and $h$. With the parameters estimated above, the tempereture should be lower than 10 K.

In summary,
we have investigated charge, valley, and spin transports in normal/ferromagnetic/normal silicene junction. We have shown that the charge, valley, and spin conductances in this junction oscillate with the length of the ferromagnetic silicene. It is also found that the current through this junction is valley and spin polarized due to the coupling between valley and spin degrees of freedom, and the valley and spin polarizations can be tuned by local application of a gate voltage. In particular, we have found \textit{fully} valley and spin polarized current by applying electric field. 
We have  also obtained the condition for observing the fully valley and spin polarized transports.

Our results would be also applicable to other valley and spin coupled systems, such as monolayers of MoS$_2$ and other group-VI dichalcogenides where low energy physics are governed by massive Dirac fermion. \cite{Xiao2}

%The author thanks S. Murakami for helpful discussion.
This work was supported by Grant-in-Aid for Young Scientists (B) (No. 23740236) and the ``Topological Quantum Phenomena" (No. 25103709) Grant-in Aid for Scientific Research on Innovative Areas from the Ministry of Education, Culture, Sports, Science and Technology (MEXT) of Japan.

\end{document}